\documentclass[10pt, a4paper, twocolumn, teaser, showabstract]{naverlabseurope}

\usepackage{lipsum}
\usepackage{multicol}
\usepackage{tikz,tkz-kiviat,pgfplots}
\usepackage{wrapfig}
\usepackage{pifont}
\usepackage{todonotes}
\usepackage{tabularx}
\usepackage{multirow}
\usepackage{adjustbox}

\graphicspath{{figures/}}

\title{SPLATE: Sparse Late Interaction Retrieval}

\correspondingauthor{thibault.formal@naverlabs.com}

\authors{Thibault Formal \authsep Stéphane Clinchant \authsep Hervé Déjean \authsep Carlos Lassance$^{\dagger \star}$}
\affiliations{Naver Labs Europe \authsep $^\dagger$Cohere}
\contributions{$^{\star}$ Work done while at Naver.}
\website{}
\websiteref{}

\begin{abstract}

  The late interaction paradigm introduced with ColBERT stands out in the neural Information Retrieval space, offering a compelling effectiveness-efficiency trade-off across many benchmarks. Efficient late interaction retrieval is based on an optimized multi-step strategy, where an approximate search first identifies a set of candidate documents to re-rank exactly. In this work, we introduce SPLATE, a simple and lightweight adaptation of the ColBERTv2 model which learns an ``MLM adapter'', mapping its \emph{frozen} token embeddings to a sparse vocabulary space with a partially learned SPLADE module. This allows us to perform the candidate generation step in late interaction pipelines with traditional sparse retrieval techniques, making it particularly appealing for running ColBERT in CPU environments. Our SPLATE ColBERTv2 pipeline achieves the same effectiveness as the PLAID ColBERTv2 engine by re-ranking 50 documents that can be retrieved under 10$ms$.

\end{abstract}

\begin{document}

\maketitle

\section{Introduction}

In the landscape of neural retrieval models based on Pre-trained Language Models (PLMs), the late interaction paradigm -- introduced with the ColBERT model~\cite{colbert} -- delivers state-of-the-art results across many benchmarks. ColBERT -- and its variants~\cite{santhanam2021colbertv2,qian2022multivector_sparse,lee2023rethinking,10.1145/3471158.3472250,10.1145/3511808.3557367,yao2022filip,lin2023finegrained,coil} -- enjoys many good properties, ranging from interpretability~\cite{formal2020white,10.1145/3539618.3591916} to robustness~\cite{10.1145/3511808.3557312,10.1007/978-3-030-99739-7_14,10.1007/978-3-031-28244-7_40,weller2023nevir}. The fine-grained \emph{interaction} mechanism, based on a token-level dense vector representation of documents and queries, alleviates the inherent limitation of single-vector models such as DPR~\cite{karpukhin-etal-2020-dense}. Due to its \emph{MaxSim} formulation, late interaction retrieval requires a dedicated multi-step search pipeline. In the meantime, Learned Sparse Retrieval~\cite{nguyen2023unified} has emerged as a new paradigm to reconcile the traditional search infrastructure with PLMs. In particular, SPLADE models~\cite{splade,formal2022distillation,formal2021splade} exhibit strong in-domain and zero-shot capabilities at a fraction of the cost of late interaction approaches -- both in terms of memory footprint and search latency~\cite{10.1145/3477495.3531833,10.1145/3539618.3591941,qiao2023optimizing,qiao2023representation}.

In this work, we draw a parallel between these two lines of works, and show how we can simply ``adapt'' ColBERTv2 \emph{frozen} representations with a light SPLADE module to effectively map queries and documents in a sparse vocabulary space. Based on this idea, we introduce SPLATE -- for {\bf SP}arse {\bf LATE} interaction -- as an alternative approximate scoring method for late interaction pipelines. Contrary to optimized engines like PLAID~\cite{plaid_22}, our method relies on traditional sparse techniques, making it particularly appealing to run ColBERT in mono-CPU environments.
\section{Related Works}

\paragraph{\bf Efficient Late Interaction Retrieval}

Late interaction retrieval is a powerful paradigm, that requires complex engineering to scale up efficiently. Specifically, it resorts to a multi-step pipeline, where an initial set of candidate documents is retrieved based on approximate scores~\cite{colbert}. While it is akin to the traditional {\it retrieve-and-rank} pipeline in IR, it still fundamentally differs in that the same (PLM) model is used for both steps\footnote{On the contrary, a standard DPR~\cite{karpukhin-etal-2020-dense} $\gg$ MonoBERT~\cite{nogueira2019bertranker} pipeline would require feeding the query {\it twice} to a PLM at inference time.}. Late interaction models offer advantages over cross-encoders because they allow for pre-computation of document representations offline, thus improving efficiency in theory. However, this comes at the cost of storing large indexes of dense term representations. Various optimizations of the ColBERT engine have thus been introduced~\cite{eaat,santhanam2021colbertv2,plaid_22,10.1145/3477495.3531835,li-etal-2023-citadel,engels2023dessert,10.1145/3511808.3557367,qian2022multivector_sparse,shrestha2023espn,ann_colbert_21,10.1007/978-3-031-56060-6_1}. ColBERTv2~\cite{santhanam2021colbertv2} refines the original ColBERT by introducing residual compression to reduce the space footprint of late interaction approaches. Yet, search speed remains a bottleneck, mostly due to the large number of candidates to re-rank exactly ($>10k$)~\cite{ann_colbert_21}. {\it Santhanam et al.} identify the major bottlenecks -- in terms of search speed -- of the vanilla ColBERTv2 pipeline, and introduce PLAID~\cite{plaid_22}, a new optimized late interaction pipeline that can largely reduce the number of candidate passages without impacting ColBERTv2's effectiveness. In particular, PLAID candidate generation is based on 
three steps that leverage centroid interaction and centroid pruning -- emulating traditional Bag-of-Words (BoW) retrieval -- as well as dedicated CUDA kernels. It reduces the large number of candidate documents to re-rank, greatly offloading subsequent steps (index lookup, decompression, and scoring).

\paragraph{\bf Hybrid Models}

Several works have identified similarities between the representations learned by different neural ranking models. For instance, UNIFIER~\cite{shen2023unifier} jointly learns dense and sparse single-vector bi-encoders by sharing intermediate transformer layers. Similarly, the BGE-M3 embedding model~\cite{chen2024bge} can perform dense, multi-vector, and sparse retrieval indifferently. 
SparseEmbed~\cite{sparse_embed_22} extends SPLADE with dense contextual embeddings -- borrowing ideas from ColBERT and COIL~\cite{coil}. SLIM~\cite{SLIM_2023} adapts ColBERT to perform late interaction on top of SPLADE-like representations -- making it fully compatible with traditional search techniques. {\it Ram et al.}~\cite{ram-etal-2023-token} show that mapping representations of a dense bi-encoder to the vocabulary space -- via the Masked Language Modeling (MLM) head -- can also be used for interpretation purposes.
\section{Method}

SPLATE is motivated by two core ideas: \begin{enumerate*}
    \item PLAID~\cite{plaid_22} draws inspiration from traditional BoW retrieval to optimize the late interaction pipeline;
    \item dense embeddings can seemingly be mapped to the vocabulary space~\cite{ram-etal-2023-token}.
\end{enumerate*} Rather than proposing a new standalone model, we show how SPLATE can be used to approximate the candidate generation step in late interaction retrieval, by bridging the gap between sparse and dense models.

\paragraph{\bf Adapting Representations} SPLATE builds on the similarities between the representations learned by sparse and dense IR models. For instance, {\it Ram et al.}~\cite{ram-etal-2023-token} show that mapping representations of a dense bi-encoder with the MLM head can produce meaningful BoW. We take one step further and hypothesize that effective sparse models can be derived -- or at least \emph{adapted} -- from \emph{frozen} embeddings of dense IR models in a SPLADE-like fashion. We, therefore, propose to ``branch'' an MLM head on top of a \emph{frozen} ColBERT model.

\paragraph{\bf SPLATE}

Given ColBERT's contextual embeddings $(h_i)_{i \in t}$ of an input query or document $t$, we can define a simple ``adapted'' MLM head, by linearly mapping \emph{transformed} representations back to the vocabulary.  
Inspired by Adapter modules~\cite{pmlr-v97-houlsby19a,pfeiffer-etal-2020-mad}, SPLATE thus simply 
adapts \emph{frozen} representations $(h_i)_{i \in t}$ by learning a simple two-layer MLP, whose output is recombined in a residual fashion before ``MLM'' vocabulary projection:
 \begin{equation}
    w_{iv} = (h_i + MLP_{\boldsymbol{\theta}}(h_i))^T E_v + b_v
\end{equation}

\noindent where $w_i$ corresponds to an unnormalized log-probability distribution over the vocabulary $\mathcal{V}$ for the token $t_i$, $E_v$ is the (Col)BERT input embedding for the token $v$ and $b_v$ is a token-level bias. The residual guarantees a near-identity initialization -- making training stable~\cite{pmlr-v97-houlsby19a}.
We can then derive sparse SPLADE vectors from these logits as follows:

\begin{equation}
w_{v}=\max_{i \in t} \log \left(1 + \text{ReLU}(w_{iv}) \right), \quad v \in \{1,...,|\mathcal{V}|\}
\label{eq:splade_max}
\end{equation}

 \begin{figure}
  \begin{center}
    \includegraphics[width=\columnwidth]{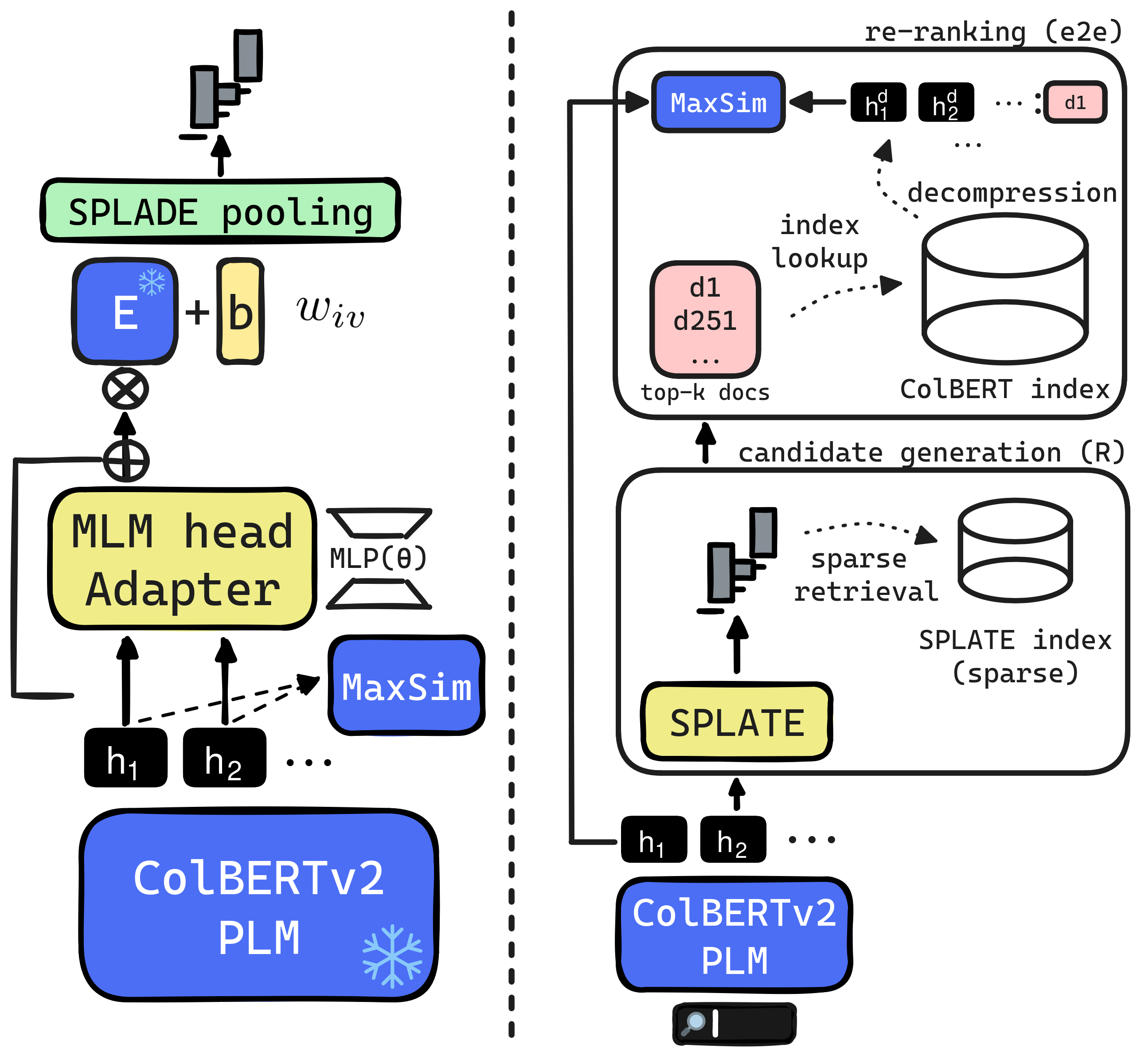}
  \end{center}
  \caption{{\it (Left)} SPLATE relies on the same representations $(h_i)_{i \in t}$ to learn sparse BoW with SPLADE (candidate generation) and to compute late interactions (re-ranking). {\it (Right)} Inference: SPLATE ColBERTv2 maps the representations of the query tokens to a sparse vector, which is used to retrieve $k$ documents from a pre-computed sparse index (\texttt{R} setting). In the \emph{e2e} setting, representations are gathered from the ColBERT index to re-rank the candidates exactly with {\it MaxSim}.}
  \label{fig:splate}
\end{figure}

We then train the parameters of the MLM head $(\boldsymbol{\theta},\boldsymbol{b})$ with distillation based on the derived SPLADE vectors to reproduce ColBERT's scores -- see Section~\ref{sec:experiments}. Our approach is very light, as the ColBERT backbone model is entirely frozen -- including the (tied) projection layer $E$. In our default setting, the MLP first down-projects representations by a factor of two, then up-projects back to the original dimension. This corresponds to a latent dimension of $768/2=384$ -- early experiments indicate that the choice of this hyperparameter is not critical -- and amounts to roughly $0.6M$ trainable parameters only (yellow blocks in Figure~\ref{fig:splate}, ({\it Left})). 

\paragraph{\bf Efficient Candidate Generation for Late Interaction}


By adapting ColBERT's frozen dense representations with a SPLADE module, SPLATE aims to approximate late interaction scoring with an efficient sparse dot product. Thus, \emph{the same representations} $(h_i)_{i \in t}$ can function in both retrieval (SPLATE module) and re-ranking (ColBERT's \emph{MaxSim}) scenarios -- \emph{requiring a single transformer inference step} on query and document sides. Thus, it becomes possible to replace the existing candidate generation step in late retrieval pipelines such as PLAID with traditional sparse retrieval to efficiently provide ColBERT with documents to re-rank. SPLATE is therefore not a model {\it per se}, but rather offers an alternative implementation to late-stage pipelines by bridging the gap between sparse and dense models. SPLATE however differs from PLAID in various aspects: \begin{itemize}
\item While PLAID implicitly derives sparse BoW representations from ColBERTv2's centroid mapping, SPLATE explicitly learns such representations by adapting a pseudo-MLM head to ColBERT frozen representations. The approximate step becomes supervised rather than (yet efficiently) ``engineered''.
\item The candidate generation can benefit from the long-standing efficiency of inverted indexes and query processing techniques such as MaxScore~\cite{10.1016/0306-45739500020-H} or WAND~\cite{10.1145/956863.956944}, making end-to-end ColBERT more ``CPU-friendly'' -- see Table~\ref{tab:tab_latency}. 
\item It is more controllable and directly amenable to all sorts of recent optimizations for learned sparse models~\cite{10.1145/3477495.3531833,10.1145/3539618.3591941}.
\item ColBERT's pipeline becomes even more interpretable, as SPLATE's candidate generation simply operates in the vocabulary space -- rather than representing documents as a lightweight bag
of centroids -- see Table~\ref{table:example} for examples. 
\end{itemize} 

Nonetheless, SPLATE requires an additional -- although light -- training round for the parameters of the Adapter module. It also requires indexing SPLATE's sparse document vectors, therefore adding a small memory footprint overhead\footnote{Note however that this is negligible compared to ColBERT's index -- for instance, the MS MARCO PISA index for the SPLATE model in Table~\ref{tab:tab_all} weighs around 2.2GB.}. Also, note that hybrid approaches like BGE-M3~\cite{chen2024bge} -- that can output sparse and multi-vector representations -- could in theory be used in late interaction pipelines. However, SPLATE is directly optimized to approximate ColBERTv2, and we leave for future work the study of jointly training the candidate generation and re-ranking modules. 
\section{Experiments}\label{sec:experiments}

\paragraph{\bf Setting} We initialize SPLATE with ColBERTv2~\cite{santhanam2021colbertv2} weights which are kept \emph{frozen}. We rely on top-$k_{q,d}$ pooling to obtain respectively query and document BoW SPLADE representations\footnote{While SPLADE is usually trained with sparse regularization, top-$k_{q,d}$ was shown to be almost as effective -- while being much simpler~\cite{nguyen2023unified}.}. We train the MLM parameters $(\boldsymbol{\theta},\boldsymbol{b})$ on the MS MARCO passage dataset~\cite{bajaj2016ms}, using both distillation and hard negative sampling. More specifically, 
we distill ColBERTv2's scores based on a weighted combination of marginMSE~\cite{hofstätter2021improving} and KLDiv~\cite{lin-etal-2021-batch} losses for $3$ epochs. We set the batch size to $24$, and select $20$ hard negatives per query -- coming from ColBERTv2's top-1000. By using ColBERTv2 as both the teacher and the source of hard negatives, SPLATE aims to approximate late interaction with sparse retrieval.
SPLATE models are trained with the SPLADE codebase on 2 Tesla V100 GPUs with 32GB memory in less than two hours\footnote{\url{https://github.com/naver/splade}}. SPLATE can be evaluated as a standalone sparse retriever (\texttt{R}), but more interestingly in an end-to-end late interaction pipeline (\texttt{e2e}) where it provides ColBERTv2 with candidates to re-rank (see Figure~\ref{fig:splate}, ({\it Right}))\footnote{Note that SPLATE (\emph{e2e}) is an alternative implementation of ColBERTv2. We use SPLATE (resp. PLAID) or SPLATE ColBERTv2 (resp. PLAID ColBERTv2) indifferently.}. For the former, we rely on the PISA engine~\cite{pisa} to conduct sparse retrieval with block-max WAND and provide latency measurements as the Mean Response Time (MRT), i.e., the average search latency measured on the MS MARCO dataset using one core of an Intel(R) Xeon(R) Gold 6338 CPU @ 2.00GHz CPU. 
For the latter, we perform on-the-fly re-ranking with the ColBERT library\footnote{\url{https://github.com/stanford-futuredata/ColBERT}}. Note that naive re-ranking with ColBERT is sub-optimal -- compared to pipelines that pre-compute document term embeddings. We leave the end-to-end latency measurements for future work -- but we believe the integration of SPLATE into ColBERT's pipelines such as PLAID should be seamless, as it would only require modifying the candidate generation step. We evaluate models on the MS MARCO dev set and the TREC DL19 queries~\cite{Craswell2019TrecDl} (in-domain), and provide out-of-domain evaluations on the 13 readily available BEIR datasets~\cite{thakur2021beir}, as well as the test pooled Search dataset of the LoTTE benchmark~\cite{santhanam2021colbertv2}.

The following experiments investigate three different Research Questions:
\begin{enumerate*}
    \item How does the sparsity of SPLATE vectors affect latency and re-ranking performance?
    \item How accurate SPLATE candidate generation is compared to ColBERTv2?
    \item How does it perform overall for in-domain and out-of-domain scenarios?
\end{enumerate*}

\paragraph{\bf Latency Results}

Table~\ref{tab:tab_latency} reports in-domain results on MS MARCO, in both retrieval-only (\texttt{R}) and end-to-end (\texttt{e2e}) settings. Overall, the results show that it is possible to ``convert'' a frozen ColBERTv2 model to an effective SPLADE, with a lightweight residual adaptation of its token embeddings. We consider several SPLATE models trained with varying pooling sizes $(k_q,k_d)$ 
-- those parameters controlling the size of the query and document representations. We observe the standard effectiveness-efficiency trade-off for SPLADE, where pooling affects both the performance and average latency. These results indicate that one can easily control the latency of the candidate generation step by selecting appropriate pooling sizes. \emph{However, after re-ranking with ColBERTv2, all the models perform comparably}, which is interesting from an efficiency perspective, as it becomes possible to use very lightweight models to cheaply provide candidates (e.g., as low as 2.9{\it ms} Mean Response Time), while achieving performance on par with the original ColBERTv2 (see Table~\ref{tab:tab_all}). For comparison, the end-to-end latency reported in PLAID \cite{plaid_22} (single CPU core, less conservative setting with $k=10$) is around 186{\it ms} on MS MARCO. Given that candidate generation accounts for around two-thirds of the complete pipeline~\cite{plaid_22}, SPLATE thus offers an interesting alternative for running ColBERT on mono-CPU environments.  

\setlength{\tabcolsep}{2pt}

\begin{table}[ht]
\centering
\caption{Retrieval latency (MRT), retrieval-only (\texttt{R}) and end-to-end (\texttt{e2e}, $k=50$) MRR@10 on MS MARCO dev.}

\adjustbox{max width=\columnwidth}{
\begin{tabular}{l c c c c}
\multirow{2}{*}{$(k_q,k_d)$}  & \multicolumn{2}{c}{\texttt{R}} & \texttt{e2e} $(k=50)$   \\
                      & MRT  ({\it ms})  &  MRR@10 &  MRR@10  \\
\midrule  
$(5,30)$  &   2.9 & 34.5 & 39.5    \\ 
$(5,50)$  &   4.3 & 35.5 &   39.7   \\ 
$(5,100)$  &  7.4 & 35.6 &   39.8   \\ 
$(10,100)$  & 24.0& 36.7 &   40.0   \\ 
$(20,200)$  &  106.0 & 37.4 &   40.0 \\ 
\end{tabular}
}
\label{tab:tab_latency}
\end{table}

\paragraph{\bf Approximation Quality}

\begin{figure}
\centering
    \centering 
    \includegraphics[width=0.45\textwidth]{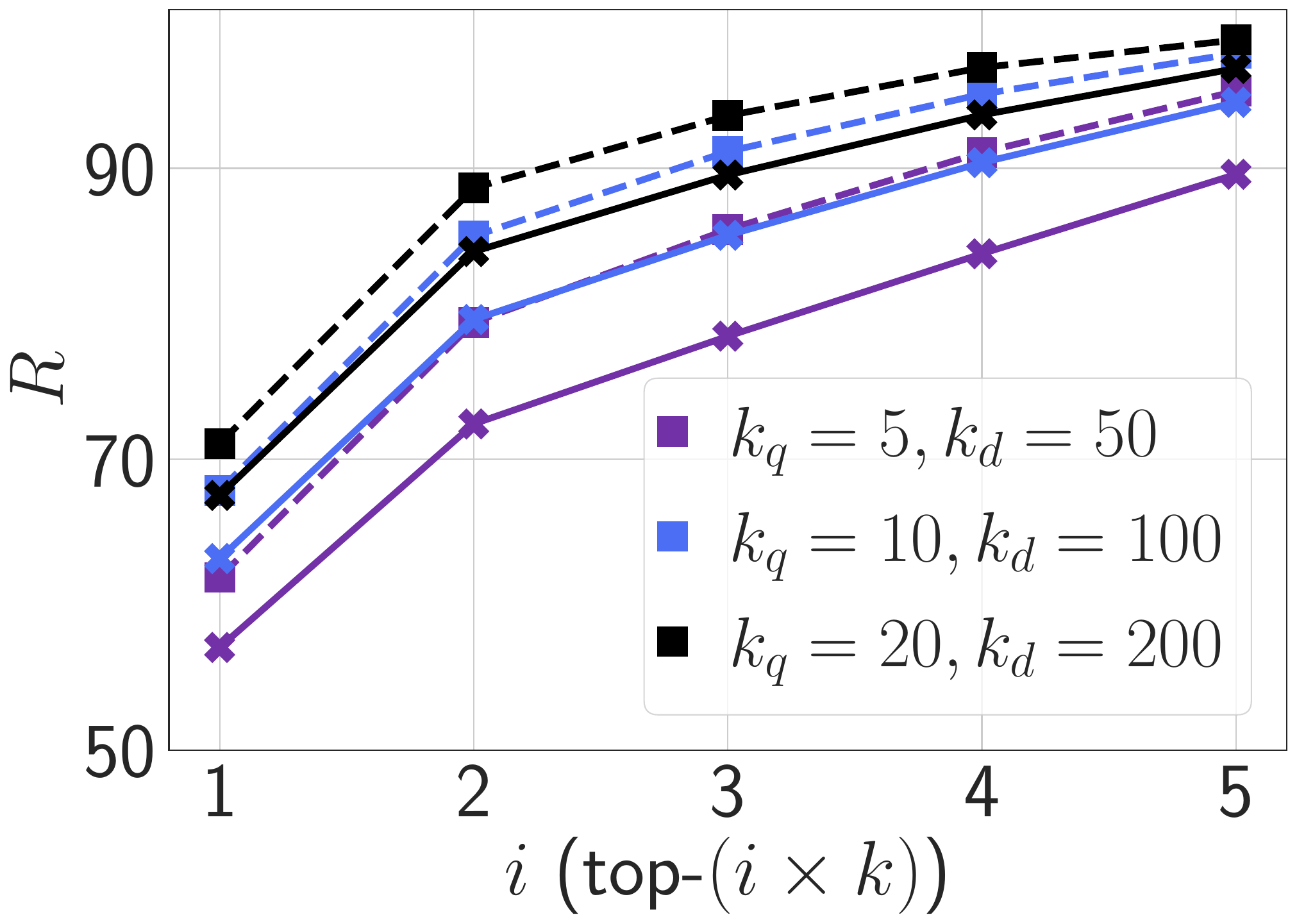}
    \caption{Candidate generation approximate accuracy on MS MARCO dev -- SPLATE (\texttt{R}). Dotted lines ($\blacksquare$) represent $R(10)$, solid lines represent (\ding{54}) $R(100)$.}
    \label{fig:a}
\end{figure}

To assess the quality of SPLATE approximation, we compare the top-$k$ passages retrieved by PLAID ColBERTv2 to the ones retrieved by SPLATE (\texttt{R}). We report in Figure~\ref{fig:a} the average fraction $R(k)$ of documents in SPLATE's top-$k'$ that also appear in the top-$k$ documents retrieved by ColBERTv2 on MS MARCO, for $k\in \{10,100\}$ and $k'=i\times k, i\in\{1,...,5\}$. When $k=10$, SPLATE can retrieve more than $90\%$ of ColBERTv2's documents in its top-50 ($i=5$), for all levels of $(k_q,k_d)$. This explains the ability of SPLATE to fully recover ColBERT's performance by re-ranking a handful of documents (e.g., 50 only). We additionally observe that the quality of approximation falls short for efficient models (i.e., lower $(k_q,k_d)$) when $k$ is higher.

\begin{figure}
\centering
    \centering 
    \includegraphics[width=0.45\textwidth]{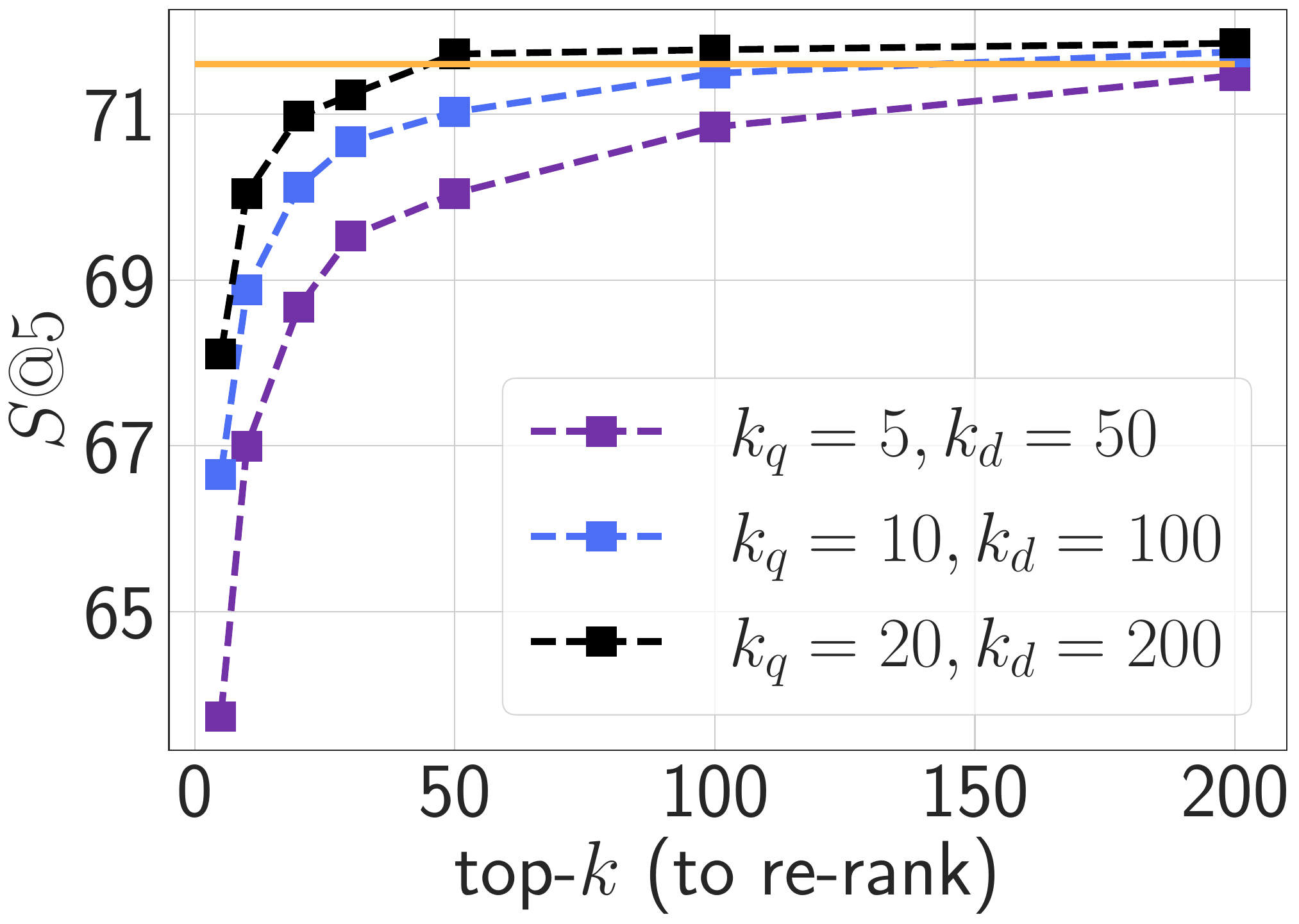}
    \caption{Impact of $k$ and $(k_q,k_d)$ on SPLATE (\texttt{e2e}) ouf-of-domain performance -- $Success@5$ on LoTTE (test pooled Search). The orange line represents ColBERTv2.}
    \label{fig:b}
\end{figure}

Figure~\ref{fig:b} further reports the performance of SPLATE (\texttt{e2e}) on out-of-domain. We observe similar trends, where increasing both the number $k$ of documents to re-rank and $(k_q,k_d)$ leads to better generalization. Overall, re-ranking only 50 documents provides a good trade-off across all settings -- echoing previous findings~\cite{plaid_22,ann_colbert_21}. Yet, the most efficient scenario ($(k_q,k_d)=(5,50)$, $k=10$) still leads to impressive results: 38.4 MRR@10 on MS MARCO dev (not shown), 70.0 $S$@5 on LoTTE (purple line on Figure~\ref{fig:b}).  

\paragraph{\bf Overall Results}

Finally, Table~\ref{tab:tab_all} compares SPLATE ColBERTv2 with the reference points ColBERTv2~\cite{santhanam2021colbertv2} and PLAID ColBERTv2 ($k=1000$)~\cite{plaid_22} -- in both \texttt{R} and \texttt{e2e} settings. We also include results from SPLADE++~\cite{formal2022distillation}, as well as the hybrid methods SparseEmbed~\cite{sparse_embed_22} and SLIM++~\cite{SLIM_2023}
-- even though they are not entirely comparable to SPLATE. While SparseEmbed and SLIM introduce new models, SPLATE rather proposes an alternative implementation to ColBERT's late retrieval pipeline. We further report the two baselines consisting of retrieving documents with BM25 (resp. SPLADE++) and re-ranking those with ColBERTv2 (BM25 $\gg$ C and S $\gg$ C respectively, with $k=50$). Note that we expect SPLATE to perform in between, as BM25 $\gg$ C relies on a less effective retriever, while S $\gg$ C fundamentally differs from SPLATE, as it is based on two different models. Specifically, it requires feeding the query to a PLM \emph{twice} at inference time. Overall, SPLATE (\texttt{R}) is effective as a standalone retriever (e.g., reaching almost 37 MRR@10 on MS MARCO dev). On the other hand, SPLATE (\texttt{e2e}) performs comparably to ColBERTv2 and PLAID on MS MARCO, BEIR, and LoTTE.
Additionally, we conducted a meta-analysis against PLAID with RANGER~\cite{sertkan2023ranger} over the 13 BEIR datasets, and found no statistical differences on 10 datasets, and statistical improvement (resp. loss) on one (resp. two) dataset(s). Finally, we provide in Table~\ref{table:example} some examples of predicted BoW for queries in MS MARCO dev -- highlighting the interpretable nature of the retrieval step in SPLATE-based ColBERT's pipeline. 

\setlength{\tabcolsep}{1pt}

\begin{table}
\normalsize
\centering
\caption{Evaluation of SPLATE with $(k_q,k_d)=(10,100)$ and $k=50$. $^{abcde}$ denote significant improvements over the corresponding rows, for a paired $t$-test with $p$-value=0.01 and Bonferroni correction (MS MARCO dev set and DL19). PLAID ColBERTv2~\cite{plaid_22} ($k=1000$) reports the \emph{dev} LoTTE$^*$ $S@5$.} 

\adjustbox{max width=\columnwidth}
{
\begin{tabular}{l c c  c c c}
\multirow{2}{*}{} & \texttt{MS MARCO} & \multicolumn{2}{c}{\texttt{DL19}} & \texttt{BEIR} & \texttt{LoTTE} \\
                      & {\footnotesize MRR@10} &  {\footnotesize nDCG@10} & 
                      {\footnotesize R@1k} &
                      {\footnotesize nDCG@10} & {\footnotesize S@5} \\
\toprule
$\blacktriangleright$ \texttt{Sparse/Hybrid} \\
SPLADE++~\cite{formal2022distillation} & 38.0 &  73.2 & 87.5 & 50.7 & - \\
SparseEmbed~\cite{sparse_embed_22} & 39.2   & - & - & 50.9 & - \\
SLIM++~\cite{SLIM_2023} & 40.4  & 71.4 & 84.2 & 49.0 & - \\
$\blacktriangleright$ \texttt{References} \\
 ColBERTv2~\cite{santhanam2021colbertv2} & $39.7$ & - & - & 49.7 & 71.6  \\
{\it (a)} PLAID ColBERTv2~\cite{plaid_22} & $39.8^{bd}$  & 74.6 & 85.2$^{b}$ & - &  69.6$^*$ \\
{\it (b)} BM25 $\gg$ C ($k=50$) &  34.3 & 68.7 &  73.9 &  49.0 & 62.8 \\
{\it (c)} S $\gg$ C ($k=50$) &  $40.4^{bd}$ &  74.4 & 87.5$^{b}$ &  49.9 & 72.0 \\
$\blacktriangleright$ \texttt{SPLATE} ColBERTv2 ($k=50$) \\
{\it (d)} SPLATE  (\texttt{R}) & $36.7^{b}$& 72.9 &  84.4$^{b}$ & 46.5 & 66.7 \\
{\it (e)} SPLATE (\texttt{e2e}) & $40.0^{bd}$ & 74.2 & 84.4$^{b}$ & $49.6$ & 71.0\\
\end{tabular}%
}
\label{tab:tab_all}
\end{table}

\begin{table}
\centering
\small
\caption{BoW SPLATE representations for queries in the MS MARCO dev set with $(k_q,k_d)=(10,100)$ (model from Table~\ref{tab:tab_all}).}
\setlength{\tabcolsep}{6pt}
\begin{tabular}{p{\columnwidth}}
\toprule
    \multicolumn{1}{c}{\textbf{SPLATE BoW}} \\
\midrule
$\mathcal{Q} \rightarrow$ ``\emph{what is the medium for an artisan}'' \\
$\blacktriangleright$ ({\it medium}, 2.2), ({\it art}, 1.8), ({\it \#\#isan}, 1.7), ({\it media}, 1.1), ({\it craftsman}, 0.9), ({\it arts}, 0.6), ({\it carpenter}, 0.6), ({\it artist}, 0.5), ({\it \#\#vre}, 0.4), ({\it draper}, 0.3) \\
\midrule
$\mathcal{Q} \rightarrow$ ``\emph{treating tension headaches without medication}'' \\
$\blacktriangleright$ ({\it headache}, 2.1), ({\it tension}, 1.8), ({\it without}, 1.6), ({\it treatment}, 1.5), ({\it treat}, 1.4), ({\it medication}, 1.3), ({\it drug}, 0.8), ({\it baker}, 0.7), ({\it no}, 0.6), ({\it stress}, 0.5) \\
\midrule
 $\mathcal{Q} \rightarrow$ ``\emph{cost of interior concrete flooring}'' \\
 $\blacktriangleright$ ({\it price}, 2.45), ({\it concrete}, 1.96), ({\it interior}, 1.85), ({\it floor}, 1.77), ({\it internal}, 1.14), ({\it \#\#ing}, 1.0), ({\it total}, 0.62), ({\it inside}, 0.57), ({\it harrison}, 0.56), ({\it cement}, 0.26)
\end{tabular}
\label{table:example}
\end{table}

To sum up, our results demonstrate that the SPLATE implementation of ColBERTv2 (i.e., SPLATE (\texttt{e2e})) can bridge the gap with the original late interaction pipelines, by re-ranking a much lower number of documents -- similar to the PLAID engine. However, the sparse term-based nature of the candidate generation step makes it particularly appealing in mono-CPU environments efficiency-wise.
\section{Conclusion}

We propose SPLATE, a new lightweight candidate generation technique simplifying ColBERTv2's candidate generation for late interaction retrieval. SPLATE adapts ColBERTv2's frozen embeddings to conduct efficient sparse retrieval with SPLADE. When evaluated end-to-end, the SPLATE implementation of ColBERTv2 performs comparably to ColBERTv2 and PLAID on several benchmarks, by re-ranking a handful of documents. Beyond optimizing late interaction retrieval, our work opens the path to a deeper study of the link between the representations trained from different architectures.

{
    \small
    \bibliographystyle{ieeenat_fullname}
    \bibliography{main}
}


\end{document}